\documentclass[a4paper,fleqn,usenatbib,useAMS]{mnras}

\usepackage{newtxtext,newtxmath}

\usepackage[T1]{fontenc}
\usepackage{ae,aecompl}
\usepackage{color}

\newcommand{\CI}{C\,{\normalsize \textit{\rm I}}}
\newcommand{\CItitle}{C\,{\Large \textit{\rm I}}}
\newcommand{\CIfig}{C\,{\sevensize {\rm I}}}
\newcommand{\BlakeNote}{$\sim$10$^{-4}$}
\newcommand{\HII}{H\,{\normalsize \textit{\rm II}}}

\usepackage{graphicx}	
\usepackage{amsmath}	
\usepackage{amssymb}	

\title[Atomic Carbon, CO, and Dust Emission in A LIRG]{The 300-pc Scale ALMA View of [\CItitle]~$^3P_1$--$^3P_0$, CO~$J$ = 1--0, and 609~$\mu$m Dust Continuum in A Luminous Infrared Galaxy}

\author[Toshiki Saito et al.]{
T. Saito,$^{1}$\thanks{E-mail: saito@mpia.de}
T. Michiyama,$^{2}$
D. Liu,$^{1}$
Y. Ao,$^{3,4}$
D. Iono,$^{2,5}$
K. Nakanishi,$^{2,5}$
\newauthor
E. Schinnerer,$^{1}$
K. Tadaki,$^{2}$
J. Ueda,$^{2}$
and
T. Yamashita$^{2}$
\\
$^{1}$Max Planck Institute for Astronomy, K\"{o}nigstuhl 17, 69117 Heidelberg, Germany\\
$^{2}$National Astronomical Observatory of Japan, 2-21-1 Osawa, Mitaka, Tokyo 181-8588, Japan\\
$^{3}$Purple Mountain Observatory, Chinese Academy of Sciences, Nanjing 210008, China\\
$^{4}$School of Astronomy and Space Science, University of Science and Technology of China, Hefei, Anhui, China\\
$^{5}$The Graduate University for Advanced Studies (SOKENDAI), 2-21-1 Osawa, Mitaka, Tokyo 181-0015, Japan
}

\date{Accepted XXX. Received YYY; in original form ZZZ}

\pubyear{2020}

\begin{document}
\label{firstpage}
\pagerange{\pageref{firstpage}--\pageref{lastpage}}
\maketitle

\begin{abstract}
We present high-quality ALMA Band~8 observations of the [\CI] $^3P_1$--$^3P_0$ line and 609~$\mu$m dust continuum emission toward the nearby luminous infrared galaxy (LIRG) IRAS~F18293-3413, as well as matched resolution (300-pc scale) Band~3 CO~ $J=$1--0 data, which allow us to assess the use of the [\CI] $^3P_1$--$^3P_0$ line as a total gas mass estimator. We find that the [\CI] line basically traces structures detected in CO (and dust), and a mean (median) [\CI]/CO luminosity ($L'_{\rm [\CI]}$/$L'_{\rm CO}$) ratio of 0.17 (0.16) with a scatter of 0.04. However, a pixel-by-pixel comparison revealed that there is a radial $L'_{\rm [\CI]}$/$L'_{\rm CO}$ gradient and a superlinear $L'_{\rm CO}$ vs. $L'_{\rm [\CI]}$ relation (slope = 1.54 $\pm$ 0.02) at this spatial scale, which can be explained by radial excitation and/or line opacity gradients.
Based on the molecular gas masses converted from the dust continuum emission, we found that the CO-to-H$_2$ and [\CI]-to-H$_2$ conversion factors are relatively flat across the molecular gas disk with a median value of 3.5$^{+1.9}_{-1.3}$ and 20.7$^{+9.2}_{-4.9}$~$M_{\odot}$ (K km s$^{-1}$ pc$^2$)$^{-1}$, respectively. A non-LTE calculation yields that typical molecular gas properties seen in nearby (U)LIRGs ($n_{\rm H_2}$ = 10$^{3-4}$~cm$^{-3}$, $T_{\rm kin}$ $\sim$ 50~K, and $X_{\rm\CI}$ = (0.8-2.3) $\times$ 10$^{-5}$) can naturally reproduce the derived [\CI]-to-H$_2$ conversion factor. However, we caution that a careful treatment of the physical gas properties is required in order to measure H$_2$ gas mass distributions in galaxies using a single [\CI]~line. Otherwise, a single [\CI]~line is not a good molecular gas estimator in a spatially resolved manner.
\end{abstract}




\section{Introduction} \label{S1}
The lowest rotational transition of carbon monoxide ($^{12}$CO $J$ = 1-0, hereafter CO~(1--0) line) is the most common total H$_2$ gas mass tracer for extragalactic objects \citep{Bolatto13}, as cold H$_2$ emission is not observable due to no permanent dipole moment.
The low critical density of CO~(1--0) ($\sim$10$^{2.5}$~cm$^{-3}$ at kinetic temperature $T_{\rm kin}$ = 40~K) and the high abundance of CO relative to H$_2$ \citep[\BlakeNote;][]{Blake87} make this line an ideal tracer for the bulk of molecular gas in galaxies.
Its rest frequency (115.27120~GHz) in the millimeter (mm) window with high atmospheric transmission promises easy access from the ground.
Empirically, a conversion from CO~(1--0) line luminosity to total H$_2$ gas mass, known as $\alpha_{\rm CO(1-0)}$, is found to be rather constant in Galactic molecular clouds assuming virial equilibrium and using the large velocity gradient approximation \citep[e.g,][]{Young84}, while nearby (ultra-)luminous infrared galaxies (LIRGs and ULIRGs) show $\sim$5 times lower values \citep[e.g.,][]{Downes98}.
These variations complicate the application of $\alpha_{\rm CO(1-0)}$ in different types of galaxies, from local to high-redshift.
Moreover, observations of the relatively low-frequency CO~(1--0) line are more challenging for high-redshift galaxies compared to higher-$J$ CO lines \citep[][and references therein]{Solomon05,Carilli13}.
Thus high-redshift studies have to rely on an assumed higher-$J$ CO to CO~(1--0) luminosity ratio in order to measure total H$_2$ gas mass.

Therefore, several other molecular gas mass tracers have been proposed and their usability has been discussed for more than two decades. For example,
the lower forbidden $^3P$ fine structure line of atomic carbon ([\CI] $^3P_1$--$^3P_0$, hereafter [\CI]~(1--0) line) has been proposed, alternative to the CO lines due to its easier accessibility in the mm window for high-redshift galaxies \citep[e.g.,][]{Barvainis97,Weiss03,Papadopoulos04b,Alaghband-Zadeh13,Bothwell17,Tadaki18,Hodge20,Valentino20}, and this line is the main target of this paper.

The rest frequency of [\CI]~(1--0) line (492.16065~GHz) allows, for example, a receiver covering 2mm and 3mm windows to detect high-redshift galaxies at z = 2.0--4.8 depending on the atmospheric transmission. As the critical density is similar to that of the CO~(1--0) line ($\sim$10$^{2.7}$~cm$^{-3}$ at $T_{\rm kin}$ = 40~K), the [\CI]~(1--0) line could be a better tracer of total H$_2$ gas mass in high-redshift systems than higher-$J$ CO lines, which may no longer trace diffuse, cold molecular gas in starburst galaxies \citep[e.g.,][]{Rangwala11}. As reported by observational studies for Galactic molecular clouds \citep[][]{Phillips81,Ojha01,Oka01,Tanaka11,Shimajiri13}, the [\CI]~(1--0) emission appears to coincide with CO~(1--0) emission at $\sim$0.1-1-pc scale resolution. These observational results are inconsistent with the classical view of photodissociation regions \citep[PDRs;][]{Hollenbach97}. In classical PDR theories, ultraviolet (UV) photon flux from young, hot stars is sufficient to control the physical properties of the cold interstellar medium (ISM), and thus the ISM properties strongly vary as a function of the distance from UV sources, forming C$^+$/C/CO layers between the diffuse and dense gas phases. However, recent numerical simulations predict that the atomic carbon abundance significantly increases throughout clouds in star-forming galaxies as the cosmic ray ionization rate is expected to be much higher (i.e., cosmic rays can penetrate into and efficiently ionize the cold ISM), and thus \CI\, coincides with CO \citep[e.g.,][]{Offner14,Glover16,Bisbas17,Papadopoulos18,Clark19}.

These studies provide observational and theoretical evidence that support the application of the [\CI]~(1--0) line as a total H$_2$ gas mass tracer similar to the CO~(1--0) line \citep[see also][]{Papadopoulos04a}. However, only very few studies in the literature used high quality maps to test the large-scale (kpc-scale) correlation between [\CI]~(1--0) and CO~(1--0) within galaxy disks, which can differ from the spatial correlation seen within molecular clouds, as most of the previous works focused on galaxy-to-galaxy variations using single-dish data \citep[e.g.,][]{Jiao17}, central regions ($<$1~kpc) of nearby galaxies \citep[e.g.,][]{Krips16,Izumi18,Miyamoto18}, and unresolved or marginally-resolved nearby and high-redshift sources \citep[e.g.,][]{Hughes17,Jiao19,Nesvadba19}.

In addition to the [\CI]~(1--0) line, the Rayleigh-Jeans part of the thermal dust continuum emission at mm and sub-mm wavelengths have started to be employed as an independent total H$_2$ gas mass tracer for high redshift galaxies based on an empirically calibrated relation between optically-thin cold dust emission and the mass of interstellar dust and gas in local star-forming galaxies, the Milky Way, and submillimeter galaxies (SMGs) \citep[e.g.,][]{Eales12,Scoville14,Groves15,Liu19}. This approach takes advantages of the higher sensitivity of mm/sub-mm facilities to continuum emission, allowing us to efficiently establish a large sample of high redshift star-forming galaxies \citep[e.g,][]{Aravena16,Hodge20}. However, the general spatially-resolved (kpc-scale) properties of CO and dust emission in extragalactic objects, especially starburst galaxies such as (U)LIRGs as a local counterpart of high redshift SMGs, are not fully understood yet \citep[e.g.,][]{Iono07,Wilson08,Sakamoto14,Saito15}.

In this paper, we present high-quality Atacama Large Millimeter/submillimeter Array (ALMA) Band~3 and 8 observations toward the disk ($\sim$4.5~kpc in diameter) of a nearby LIRG at $\sim$300~pc resolution.
The 300~pc resolution matches to one of the largest molecular gas structures seen in merging galaxies (e.g., NGC~4038/9), giant molecular associations \citep[a few 100 pc;][]{Wilson00,Ueda12}.
Our goals are to provide a direct comparison among [\CI]~(1--0), CO~(1--0), and thermal dust continuum emission within the gas disk of a galaxy with well below kpc resolution and sufficient sensitivity and to test whether the [\CI]~(1--0) line can be used as a total H$_2$ gas mass tracer for spatially-resolved studies on extragalactic objects or not.

Our target, IRAS~F18293-3413, is one of the nearby LIRGs classified as an \HII\, galaxy \citep[$D_{\rm L}$ = 86.0~Mpc and $L_{\rm IR}$ = 10$^{11.88}$~$L_{\odot}$;][]{Armus09}.
It is a mid-to-late stage merger with an elliptical companion galaxy at a projected distance of $\sim$4.5~kpc.
Several hard X-ray works suggested that there is no strong AGN contributing to the X-ray spectrum \citep[e.g.,][]{Risaliti20}.
This is consistent with the multi-wavelength study done by \citet{Herrero-Illana17}, which concluded that a possible AGN contribution to the total luminosity is less than 0.2\%.
The total star formation rate (SFR) derived from Paschen $\alpha$ emission and spectral energy distribution (SED) fitting is $\sim$100~$M_{\odot}$ yr$^{-1}$ \citep{Tateuchi15,Shangguan19}.
The SED fitting yields that the total stellar mass is $\sim$10$^{10.94}$~$M_{\odot}$, implying $\sim$2 dex higher log (SFR) compared to the local main-sequence galaxies \citep{Saintonge16}.
The apparent brightness, thanks to the extremely high SFR, and the relatively face-on molecular gas disk, makes it a primary target to study the [\CI], CO, and dust continuum emission with high signal-to-noise ratios.

\begin{figure*}
\includegraphics[width=2\columnwidth]{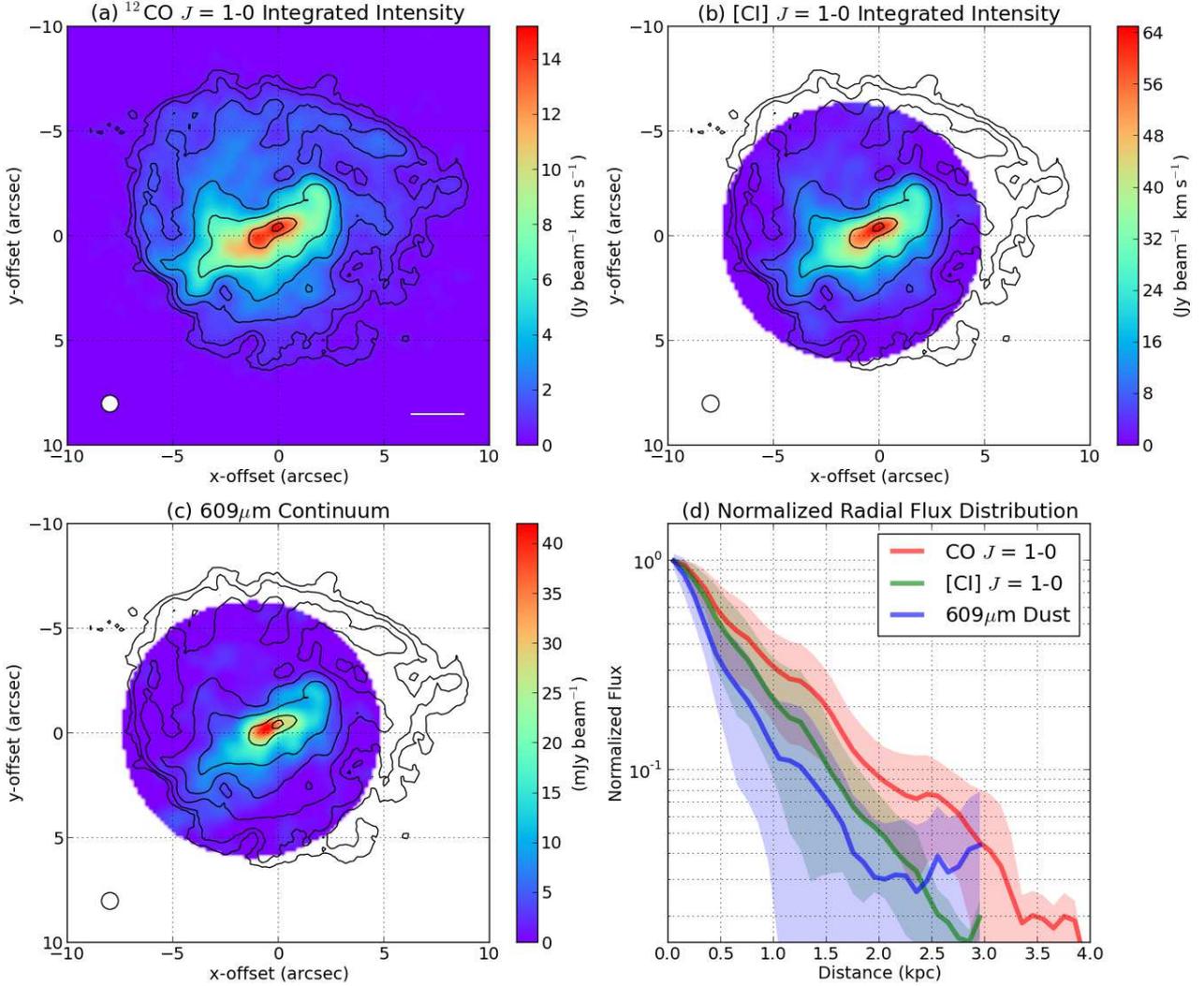}
\caption{
(a) The CO~(1--0) integrated intensity image of IRAS~F18293-3413. The contours are (0.025, 0.05, 0.1, 0.2, 0.4, 0.8, and 0.95) $\times$ 15.2~Jy beam$^{-1}$ km s$^{-1}$. The convolved synthesized beam size is shown in the lower left corner.
The scale bar in the lower right corner corresponds to 1~kpc.
(b) The [\CIfig]~(1--0) integrated intensity image. The contours are the same as in panel (a).
(c) The 609~$\mu$m continuum image. The contours are the same as in panel (a). Note that all images shown here are after correcting for primary beam attenuation.
(d)
Azimuthally-averaged, normalized radial flux distributions of the CO~(1--0), [\CIfig]~(1--0) and 609~$\mu$m continuum images of IRAS~F18293-3413. The shaded areas show the standard deviation of the distributions.
The extent of the [\CIfig] and dust profiles are limited to the Band~8 FoV (= half power beamwidth).
}
\label{fig01}
\end{figure*}

This paper is organized as follows. We briefly summarize the ALMA data and the procedure of data reduction in Section~\ref{S2}. Section~\ref{S3} provides image properties via pixel-by-pixel comparison. Then, we discuss the capability of the [\CI]~(1--0) line to trace total H$_2$ gas mass in Section~\ref{S4}, and briefly summarize out findings in Section~\ref{S5}. We have adopted H$_0$ = 70 km s$^{-1}$ Mpc$^{-1}$, $\Omega_m$ = 0.3, and $\Omega_{\Delta}$ = 0.7 throughout this paper.

\section{ALMA Observations} \label{S2}
The ALMA 12m and Atacama Compact Array 7m observations of Band~8 [\CI]~(1-0) toward IRAS~F18293-3413 were carried out for ALMA cycle 3 and 6 programs, respectively (ID: 2015.1.01191.S; PI =  Z. Zhang, and 2018.1.00994.S; PI = T. Michiyama). These observations were tuned to the redshifted [\CI]~(1-0) emission line ($\nu_{\rm obs}$ = 483.37293~GHz).

The ACA 7m Band~8 data were obtained on 2019 May 24th using eleven 7m antennas with a projected baseline length of 8.5-47.0~m and an on-source integration time of 5.03~minutes. The 7m correlator was configured to have four spectral windows (SPWs), two of which were set to each sideband, and each of the SPWs with a 2.000~GHz bandwidth and 15.625~MHz resolution ($\sim$9.7~km s$^{-1}$). The bright quasar J1924-2914 was observed as flux and bandpass calibrators. J1802-3940 and J1733-3722 were observed as phase calibrator and check source, respectively.

The ALMA 12m Band~8 data were obtained on 2016 March 28th using forty-two 12m antennas with a projected baseline length of 14-450.0~m and an on-source integration time of 11.32 minutes. The 12m correlator was also configured to have four SPWs. Two of them were set to each sideband, and three of them have a 2.000~GHz bandwidth and 15.625~MHz resolution ($\sim$9.7~km s$^{-1}$), whereas the another one containing the [C${\rm I}$] line has a 1.875~GHz bandwidth with 0.488~MHz resolution. Pallas, J1924-2914, J1802-3940, and J1826-3650 were observed as flux calibrator, bandpass calibrator, phase calibrator, and check source, respectively.

The ALMA 12m Band~3 data were obtained on 2016 July 25th as a part of the Cycle 3 data described above. Fourty-four 12m antennas were used to cover projected baselines of 12.3-1076.1~m length. The Band~3 receivers were tuned to cover the redshifted CO~(1--0) line ($\nu_{\rm obs}$ = 113.21299~GHz). Two of four SPWs were set to each sideband, and all SPWs have a 1.875~GHz with 0.977~MHz ($\sim$2.6~km s$^{-1}$) resolution. Pallas, J1924-2914, J1826-3650 are used as a flux, bandpass, and phase calibrator.

We performed calibration and imaging ({\tt tclean}) using the software {\tt CASA} \citep[version 4.5.2 and 5.4.0;][]{McMullin07}. We ran the observatory-provided calibration pipeline with a few manual data flagging. Images were reconstructed with a natural and Briggs (robust = 0.5) weighting for the Band 8 data (i.e., \CI\, and continuum) and Band 3 data (i.e., CO), respectively. We made datacubes with a velocity resolution of 10~km s$^{-1}$. Continuum emission was subtracted in the $uv$-plane by fitting the line-free channels in all available SPWs with a first order polynomial function. The line-free channels were used to create a continuum image using the multi-frequency synthesis method (mfs). After the imaging procedure, all datacubes and continuum image were convolved to 0\farcs8 resolution ($\sim$300~pc) for the sake of simplicity, which is slightly larger than the major axis of the clean beam of the CO~(1-0) data. Then, we resampled the data onto a grid with 0\farcs18 pixel size, corresponding to oversampling by a factor of $\simeq$4.53 ($\simeq$ $\pi$/$\ln 2$).

The noise rms per channel of the CO~(1--0) datacube is 2.5 mJy beam$^{-1}$, which is better than that of the [\CI]~(1--0) datacube (5.5 mJy beam$^{-1}$). The continuum data has a noise rms of 0.75 mJy beam$^{-1}$. The systematic errors on the absolute flux calibration using a solar system object are estimated to be 5\% and 20\% for both sidebands in Band~3 and Band~8, respectively.

In order to create integrated intensity maps, we first convolved the CO datacube to a round 1\farcs0 resolution, then create a signal mask (i.e., True/False mask) with pixels larger than 3$\sigma$ level. We repeated the procedure for a datacube convolved to 3\farcs5 resolution, and applied a 5$\sigma$ level. Next, we created a combined mask using pixels where both masks are True. This masking procedure allows us to pick strong detections with surrounding weak emission which are hard to see in the original datacubes, and at the same time, we avoid inclusion of faint isolated patchy structures which are typical spurious structures. Finally, we collapsed the original 0\farcs8 resolution datacubes by applying the combined mask and 2$\sigma$ threshold. The products are shown in Figure~\ref{fig01}.

\section{Results} \label{S3}
\subsection{Spatial Distributions}
In Figure~\ref{fig01}a-c, we show the 0\farcs8 ($\sim$300~pc) resolution moment-0 maps of the CO~(1--0) and [\CI]~(1--0) lines plus the 609~$\mu$m continuum map. The [\CI]~(1--0) line and a continuum emission basically trace well structures present in CO~(1--0) line emission, simply implying that the [\CI] line as well as the cold dust continuum can be used as a tracer of extragalactic molecular gas mass.

The total integrated flux of [\CI]~(1--0) is 1550 $\pm$ 310 Jy km s$^{-1}$. The total integrated flux of CO~(1--0) inside the field of view (FoV) of the Band~8 data ($\sim$13\farcs0 in full width at half maximum) is 472 $\pm$ 24 Jy km s$^{-1}$, which is 87\% of the total integrated flux of the CO~(1--0) map inside the Band~3 FoV (541 $\pm$ 27 Jy km s$^{-1}$). This implies that, assuming as an extreme case that [\CI]~(1--0) emission traces similar structures seen in CO~(1--0) emission, the [\CI]~(1--0) total flux inside the Band~3 FoV should be 1780 $\pm$ 356 Jy km s$^{-1}$. We regard this value as the [\CI]~(1--0) total flux of our data in order to estimate the amount of missing flux. The [\CI]~(1--0) {\it Herschel}/SPIRE single-dish flux is 2300 $\pm$ 370 Jy km s$^{-1}$ \citep{Kamenetzky16}, for which we assumed a systematic flux uncertainty of 16\% \citep{Rosenberg15}. This indicates that the recovered [\CI]~(1--0) flux is 77 $\pm$ 20 \%, which is consistent with the recovered CO~(1--0) flux of 79 $\pm$ 9 \% \citep[NRAO~12m flux = 686 $\pm$ 69 Jy km s$^{-1}$; ][]{Garay93}.
If we assume the opposite extreme case that the SPIRE [\CI]~(1--0) total flux comes from an area within the Band~8 FoV of $\sim$13\farcs0, the recovered [\CI]~(1--0) flux will be 67 $\pm$ 17 \%.
This is on the order of the absolute flux calibration uncertainties between the two observations. Thus, throughout this paper, we do not consider that the missing flux affects our analysis and conclusion. We note that the step calculating the expected total [\CI]~(1--0) flux inside the Band~3 FoV carries a significant uncertainty.

We show the radial flux distributions of CO~(1--0), [\CI]~(1--0), and 609$\mu$m continuum in Figure~\ref{fig01}d. We define the peak position of the CO~(1--0) emission as the centre. We found that the apparent disk size is different among the gas mass tracers (CO $>$ [\CI] $>$ dust). Recent high-resolution CO and dust observations toward the nearby merging (U)LIRGs have revealed that 850~$\mu$m dust continuum distribution is typically more compact than the low-$J$ CO flux distribution in nearby (U)LIRGs \citep[e.g.,][]{Wilson08,Sakamoto14,Saito15,Saito17}. Those differences may be driven by a combination of temperature, optical depth, and abundance gradients, which can be revealed through multiple frequency and line observations.

\begin{figure*}
\includegraphics[width=2\columnwidth]{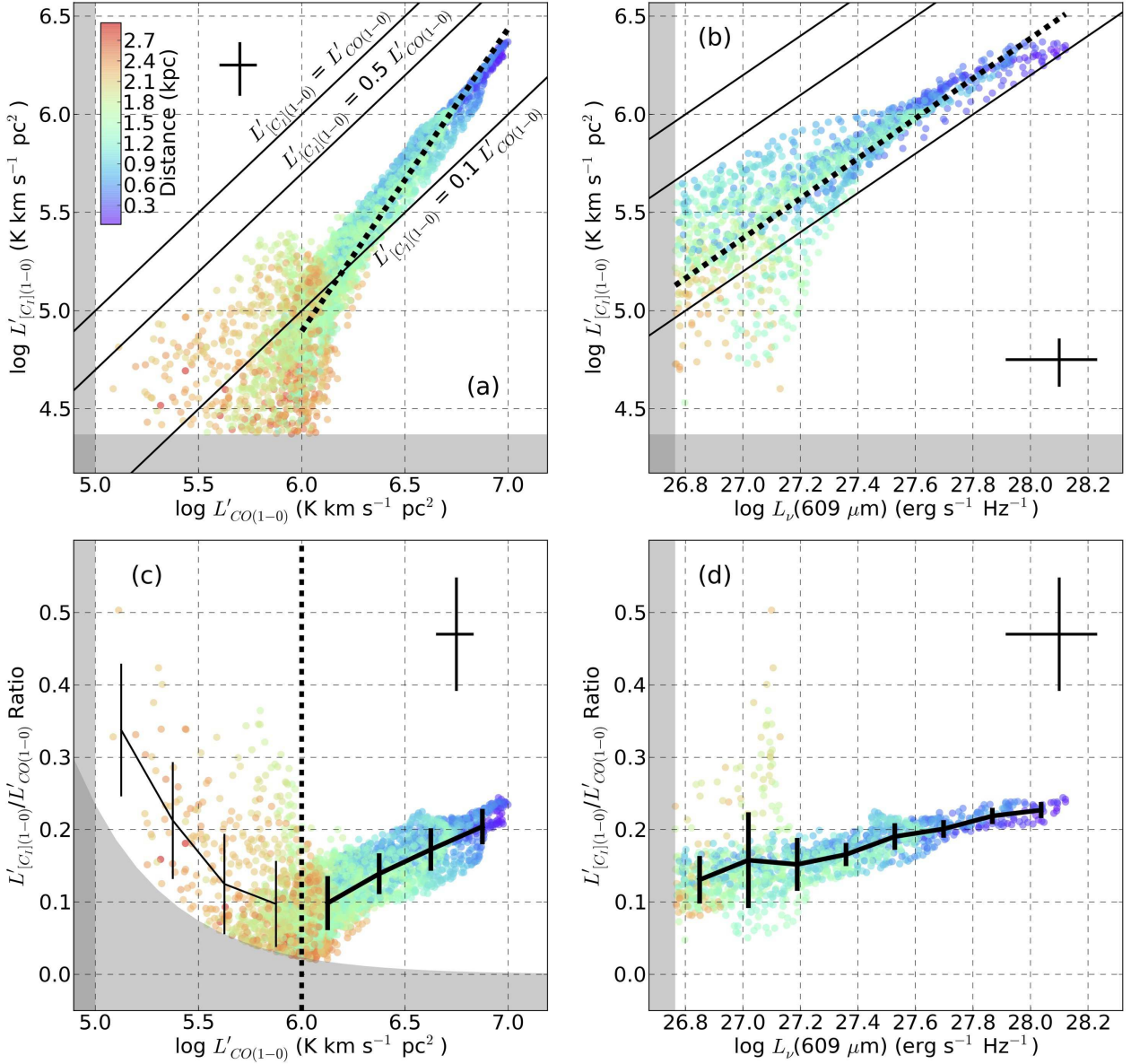}
\caption{
(a) Pixel-by-pixel comparison between [\CIfig]~(1--0) and CO~(1--0) luminosities toward IRAS~F18293-3413. The color scale corresponds to distance from centre. Typical error bars including the statistical error and the systematic flux error are shown in the top left corner. Grey areas have less than 3$\sigma$ sensitivity. Three black solid lines indicate linear relations for [\CIfig]/CO = 1.0, 0.5, and 0.1. The dashed black line shows the best fit relation to the data points with log ($L'_{\rm CO(1-0)}$ (K km s$^{-1}$ pc$^2$)$^{-1}$) $>$ 6.0 (see text).
(b) Pixel-by-pixel comparison between [\CIfig]~(1--0) and 609~$\mu$m luminosities. The three black solid lines are the same as those in panel (a) when log $L_{\nu}$(609~$\mu$m) = 20.8 + log $L'_{\rm CO(1-0)}$.
The dashed black line shows the best fit relation to the data points.
(c) Pixel-by-pixel comparison between [\CIfig]-to-CO luminosity ratio and CO~(1--0) luminosity. A dashed vertical line presents the completeness limit of log ($L'_{\rm CO(1-0)}$ (K km s$^{-1}$ pc$^2$)$^{-1}$) = 6.0. Binned distributions showing the median with the standard deviation at each bin are shown as black solid lines.
The data points are divided into 8 equally-spaced bins.
(d) Pixel-by-pixel comparison between [\CIfig]-to-CO luminosity ratio and 609~$\mu$m luminosity.
The very high ratios seen in the lower luminosity regimes mainly come from pixels which satisfies (1) low signal-to-noise CO~(1--0) and (2) located at the edge of the Band 8 FoV.
}
\label{fig02}
\end{figure*}

\begin{figure}
\includegraphics[width=\columnwidth]{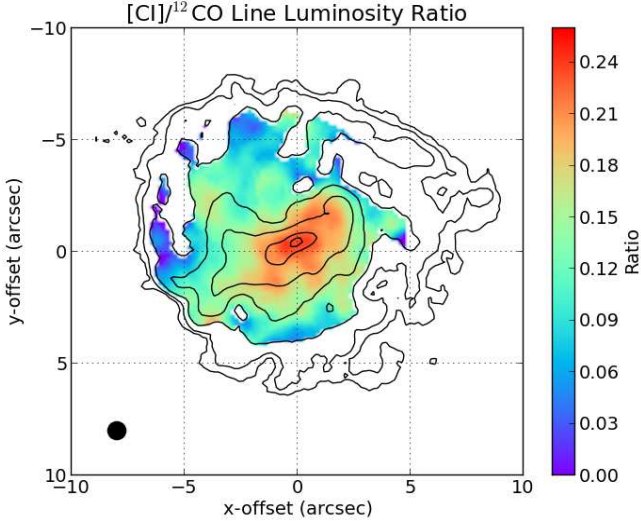}
\caption{
A map of the [\CIfig]~(1--0)/CO~(1--0) line luminosity ratio of IRAS~F18293-3413. All pixels lower than log $L'_{\rm CO(1-0)}$ = 6.0~K km s$^{-1}$ pc$^2$ are clipped (see text for details). The convolved synthesized beam is shown in the lower left corner. The contours are the same as in Figure~\ref{fig01}a.
}
\label{fig03}
\end{figure}

\subsection{Pixel-by-pixel Comparison}
We calculate luminosities of both lines ($L'_{\rm line}$) in units of K km s$^{-1}$ pc$^2$ by using the following equation \citep{Solomon05};

\begin{equation} \label{eqn1}
L'_{\rm line} = 3.25 \times 10^7 S_{\rm line}\Delta {\rm v}\:\nu_{\rm obs}^{-2}D_{\rm L}^2(1 + z)^{-3}
\end{equation}
where $S_{\rm line}\Delta$v is the velocity-integrated flux of a line in Jy km s$^{-1}$, $\nu_{\rm obs}$ is the observed frequency of the line in GHz, $D_{\rm L}$ is the source luminosity distance in Mpc, and z is the source redshift. The continuum luminosity at the frequency of $\nu$ ($L_{\nu}$) in units of erg s$^{-1}$ Hz$^{-1}$ is calculated via;
\begin{equation}
L_{\nu} = 1.20 \times 10^{27}S_{\nu_{\rm obs}}D_{\rm L}^2(1 + z)^{-3}
\end{equation}
where $S_{\nu}$ is the observed continuum flux density in Jy.

Pixel-by-pixel comparisons between CO and [\CI] luminosities and between 609~$\mu$m dust continuum and [\CI] luminosities are shown in Figures~\ref{fig02}a and \ref{fig02}b, respectively, both of which show a clear correlation. We would like to remind the readers that the pixel size is 0\farcs18 ($\sim$65~pc) and correlates across the beam ($\sim$300~pc). Although our sensitivity limits (grey shaded areas) look sufficient to evaluate the correlation strength, the scatter seems to hit the [\CI] sensitivity limit (Figures~\ref{fig02}a).

To further investigate this, we compare the [\CI]/CO luminosity ratio against the CO luminosity pixel-by-pixel (Figure~\ref{fig02}c). Now it becomes apparent that the current [\CI] sensitivity is limiting the completeness at log ($L'_{\rm CO(1-0)}$/(K km s$^{-1}$ pc$^2$)) $<$ 6.0. This is also seen in the turnover of the binned ratio distribution around log ($L'_{\rm CO(1-0)}$/(K km s$^{-1}$ pc$^2$)) = 6.0. Thus, when comparing [\CI] with CO, we decided to use data points with high completeness hereafter (i.e., log ($L'_{\rm CO(1-0)}$/(K km s$^{-1}$ pc$^2$)) $>$ 6.0), which biases our results to brighter pixels, although we are able to use $\gtrsim$80\% of pixels inside the Band 8 FoV. Deeper [\CI] data is required to overcome this limitation.
In the case of the [\CI]/CO luminosity ratio against 609~$\mu$m dust continuum luminosity (Figure~\ref{fig02}d), we do not consider the completeness limit described above as one of the advantages of this comparison is to remove the correlation between the x and y axes, though the limited [CI] sensitivity is still an issue especially the lower luminosity regime.

\subsubsection{[\CI]~(1--0) vs. CO~(1--0)}

We carried out a linear fitting to the [\CI]~(1--0) and CO~(1--0) data points with high completeness in Figure~\ref{fig02}a, and we found a superlinear relation;
\begin{equation}
\log_{10}\:L'_{\rm [C_I](1-0)} = (1.54\pm0.02)\:\log_{10}\:L'_{\rm CO(1-0)} - (4.35\pm0.15).
\end{equation}
This is also seen in the positive slope in Figure~\ref{fig02}c.

In Figure~\ref{fig03}, we show a map of the [\CI]/CO line luminosity ratio map above the completeness limit. This is the first 300-pc scale view of the [\CI]/CO line ratio toward the whole gas disk of a LIRG. There is a tendency that the central region has higher ratio ($\sim$0.24) and the outer parts of the galaxy show lower values ($\sim$0.1). The ratio variation in IRAS~F18293-3413 is within the variation of single-dish global values found for extragalactic objects \citep[0.2 $\pm$ 0.2; ][]{Gerin00,Jiao17}.

\begin{figure*}
\includegraphics[width=2\columnwidth]{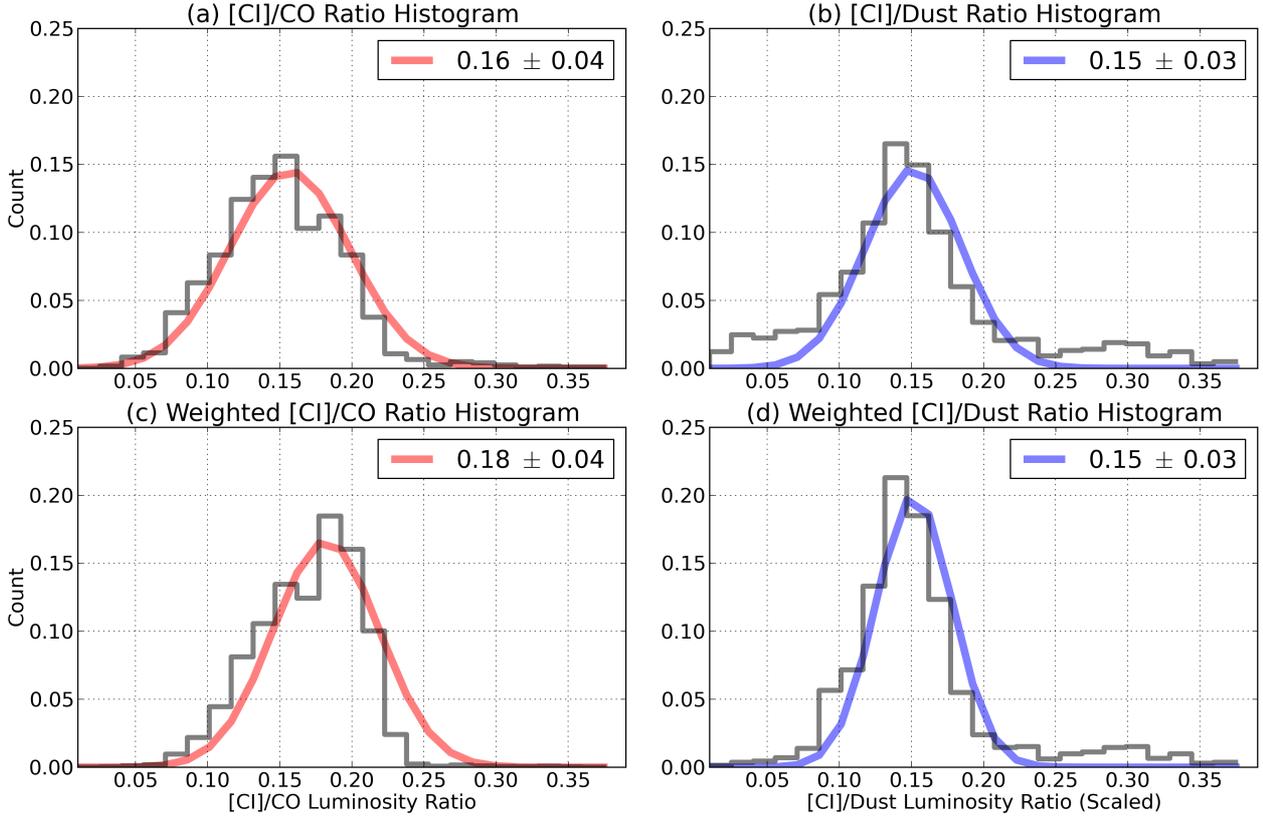}
\caption{
(a) Histogram of the [\CIfig]~(1--0)/CO~(1--0) line luminosity ratio. A best-fit Gaussian fit to the histogram is shown by the coloured bins.
(b) Histogram of the [\CIfig]~(1--0)/dust continuum luminosity ratio scaled by 10$^{20.8}$ assuming log $L_{\nu}$(609~$\mu$m) = 20.8 + log $L'_{\rm CO(1-0)}$.
(c/d) Same as panel (a/b), but histograms are weighted by the [\CIfig]~(1--0) luminosity.
}
\label{fig04}
\end{figure*}

We show the line luminosity ratio histogram and a best-fit Gaussian to the distribution in Figure~\ref{fig04}a. The ratio distribution is well represented by a single Gaussian with $\mu$ = 0.16 and $\sigma$ = 0.04. When the distribution is weighted by the [\CI]~(1--0) luminosity, i.e., a histogram showing total [\CI]~(1--0) luminosity at a given [\CI]/CO ratio bin, the shape remains similar to a single Gaussian with th peak slightly moving to the right (Figure~\ref{fig04}c; $\mu$ = 0.18 and $\sigma$ = 0.04). This indicates that the [\CI]~(1--0) intensity mainly come from higher ratio pixels, which is consistent with the impression of Figure~\ref{fig03} what the bright centre shows higher ratios.

A similar radial trend was reported toward the circumnuclear disk of the nearby Seyfert galaxy NGC~1808 \citep{Salak19} and the central kpc of the nearby starburst galaxy NGC~253 \citep{Krips16} using well-resolved datasets with $\lesssim$100~pc resolution. They basically found regions with brighter CO emission tend to show higher [\CI]~(1--0)/CO~(1--0) ratios, that is, a superlinear relation in the  [\CI]~(1--0) and CO~(1--0) scatter plot as we found in IRAS~F18293-3413.

The supearlinear relation seen in our 300-pc scale measurement, as well as by \citet{Krips16} and \citet{Salak19}, seems inconsistent with the moderately spatially-resolved measurements of nearby normal galaxies and the global measurements of (U)LIRGs, both of which exhibit a linear relationship \citep{Jiao19}. However, \citet{Jiao19} fitted the (U)LIRG data points independently of the data points for nearby normal galaxies using a fixed slope of unity (dashed-dotted blue line of their Figure~3), which is consistent with their earlier work which presented an almost linear fit of their (U)LIRG sample with a free slope \citep{Jiao17}. They did the same fitting for the nearby galaxy data points (dashed-dotted red line of \citealt{Jiao19}), where the relation is offset to lower [\CI]/CO ratios compared to the fitted linear relation for (U)LIRGs. This implies if one simultaneously fits all their data points with a free slope, the derived slope will be larger than unity. Thus, the superlinear relation we derived for IRAS~F18293-3413 is consistent with what \citet{Jiao19} found.


We note that the integrated luminosities of IRAS~F18293-3413 within the Band 8 FoV ($L'_{\rm CO(1-0)}$ $\sim$ 10$^{9.8}$ K km s$^{-1}$ pc$^2$ and $L'_{\rm [\CI](1-0)}$ $\sim$ 10$^{9.0}$ K km s$^{-1}$ pc$^2$) are consistent within the scatter of the CO~(1--0) vs. [\CI]~(1--0) luminosity plot for nearby (U)LIRGs \citep{Jiao17}.

\subsubsection{[\CI]~(1--0) vs. Cold Dust}

We fit the \CI\, and 609~$\mu$m dust data points (Figure~\ref{fig02}b) in the same way as described in the previous section, and found a nearly linear relation in contrast to the [\CI]~(1--0) and CO~(1--0) relation;
\begin{equation}
\log_{10}\:L'_{\rm [C_I](1-0)} = (1.02\pm0.02)\:\log_{10}\:L_{\rm \nu}(609 \mu m) - (22.12\pm0.58).
\end{equation}
The line luminosities correlate well with the dust continuum luminosity at brighter pixels, although the scatter becomes larger at the fainter dust luminosity regime. We show a comparison between the [\CI]/CO luminosity ratio against the dust continuum luminosity in Figure~\ref{fig02}d, and found a position correlation similar to that in Figure~\ref{fig02}c.

We plot [\CI]/dust luminosity ratio histograms in Figures~\ref{fig04}b and \ref{fig04}d. The x-axis is scaled by 10$^{20.8}$ to make a comparison with [\CI]/CO ratio histograms easier. The [\CI]/dust ratio histograms ($\mu$ = 0.15 and $\sigma$ = 0.03) have a similar shape to (or are very slightly narrower than) the [\CI]/CO ratio histograms. However, in contrast to the [\CI]/CO ratio, the weighted [\CI]/dust histogram gives similar the $\mu$ and $\sigma$ as the unweighted histogram, implying that both the bright centre and the fainter outer region show a similar [\CI]/dust ratio in general. The extreme values ($>$0.25 or $<$0.07) come from a small number of low signal-to-noise pixels.

In summary, we find that the [\CI]~(1--0) line is surprisingly well correlated with the 609$\mu$m dust continuum, but much less tracing the CO~(1--0) line at 300~pc scale in this galaxy.

\section{Discussion} \label{S4}
Here we discuss the origin of the slope of the correlations between different gas mass tracers presented in Section~\ref{S3}. Then, we derive the line luminosity to H$_2$ gas mass conversion factor ($\alpha_{\rm line}$) for CO~(1--0) and [\CI]~(1--0) using gas masses based on the 609$\mu$m dust continuum. Finally, we discuss which physical parameters are required to reproduce the derived $\alpha_{\rm [\CI](1-0)}$ based on non local thermodynamic equilibrium (non-LTE) radiative transfer modeling.

\subsection{Origin of the Slopes} \label{S4.1}
An observed luminosity ratio between two emission lines is a result of the combination of line opacity and gas physical properties. According to the radiative transfer equation, the observed line intensity in units of brightness temperature ($T_{\rm line}$) can be expressed as,
\begin{equation}
T_{\rm line} = \eta_{\rm bf}\left(\frac{h\nu_{\rm obs}/k}{e^{h\nu_{\rm obs}/kT_{\rm ex}} - 1} - \frac{h\nu_{\rm obs}/k}{e^{h\nu_{\rm obs}/kT_{\rm cmb}} - 1}\right)(1 - e^{-\tau_{\rm line}}),
\end{equation}
where $\eta_{\rm bf}$ is the beam filling factor, $h$ is Planck's constant, $k$ is Boltzmann's constant, $T_{\rm ex}$ is the excitation temperature of the line, $T_{\rm cmb}$ is the cosmic microwave background temperature (= 2.73~K), and $\tau_{\rm line}$ is the line opacity.

If we assume two lines (i.e., CO~(1--0) and [\CI]~(1--0) lines) have a very similar or the exactly same beam filling factor ($\eta_{\rm bf,CO(1-0)}$ $\simeq$ $\eta_{\rm bf,[\CI](1-0)}$), the line luminosity ratio ($R$) becomes,
\begin{equation} \label{eq6}
R = \frac{J(T_{\rm ex,[\CI](1-0)}) - J(T_{\rm cmb})}{J(T_{\rm ex,CO(1-0)}) - J(T_{\rm cmb})}\frac{1 - e^{-\tau_{\rm [\CI](1-0)}}}{1 - e^{-\tau_{\rm CO(1-0)}}},
\end{equation}
where $J$ = ($h\nu_{\rm obs}/k$)($e^{h\nu_{\rm obs}/kT}$ - 1). This is the same equation presented in \citet{Salak19}.

According to the equation, there are three different ways to increase $R$: (1) increase temperature when $T_{\rm ex,CO(1-0)}$ $\simeq$ $T_{\rm ex,[\CI](1-0)}$, (2) increase $\tau_{\rm [\CI](1-0)}$ when $\tau_{\rm CO(1-0)}$ $\gg$ 1, and/or (3) decrease $\tau_{\rm CO(1-0)}$ when $\tau_{\rm [\CI](1-0)}$ $\ll$ 1.
\begin{enumerate}
\item[(1)] Based on the non-LTE radiative transfer calculation done by \citet{Salak19}, the assumption that both lines are nearly thermalized is reasonable ($T_{\rm ex,CO(1-0)}$ $\simeq$ $T_{\rm ex,[\CI](1-0)}$ $\simeq$ $T_{\rm kin}$, where $T_{\rm kin}$ is the gas kinetic temperature) under typical molecular gas conditions found in nearby (U)LIRGs \citep[e.g.,][]{Saito17,Sliwa17}, i.e., relatively warm and moderately dense condition ($T_{\rm kin}$ $\sim$ 40-80~K and $n_{\rm H_2}$ $\sim$ 10$^{3-4}$~cm$^{-3}$). In this case, $R$ monotonically increases as $T_{\rm kin}$ increases. Since the central part of galaxies typically shows a stronger interstellar radiation field that is efficiently heating the surrounding material, the expected radial $T$ gradient results in a radial $R$ gradient and thus a superliear relation. This scenario is supported by \citet{Jiao19}, which show that a marginally resolved radial [\CI]~(2--1)/[\CI](1--0) gradient in the nearby edge-on starburst galaxy M82.
\item[(2)] If $T_{\rm kin}$ is almost constant (or has a negligible radial gradient) across the molecular gas disk of IRAS~F18293-3413, the observed radial $R$ gradient can be only reproduced by varying the line opacities. At first, we assume $\tau_{\rm CO(1-0)}$ $\gg$ 1. In order to reproduce the observed radial $R$ gradient from 0.25 to 0.1, $\tau_{[\CI](1-0)}$ should be moderate to optically thin across the disk (between 0.4 and 0.1).
This is consistent with the fact that [\CI]~(1--0) linearly correlates with optically-thin dust continuum emission assuming $T_{\rm ex,[\CI](1-0)}$ $\simeq$ $T_{\rm dust}$.
Note that many previous extragalactic works also argued that the [\CI]~(1--0) line is optically-thin to moderately thin based on radiative transfer analysis \citep[e.g.,][]{Israel15,Krips16}.
\item[(3)] The another possibility is a variation of $\tau_{\rm CO(1-0)}$. If $\tau_{\rm [\CI](1-0)}$ is optically-thin and almost constant radially, $\tau_{\rm CO(1-0)}$ $\lesssim$ 1 around the centre of IRAS~F18293-3413 can explain the superlinear relation. However, galaxy centres, especially (U)LIRG nuclei, typically exhibit higher H$_2$ column density, implying that $\tau_{\rm CO(1-0)}$ $\lesssim$ 1 at the centre may be unrealistic.
\end{enumerate}

We cannot reject any of these three possibilities for now, mainly because of the lack of an independent optically-thin molecular gas mass tracer (e.g., $^{13}$CO lines) and constraints on the excitation condition of the CO and [\CI] lines at the same high-resolution scale.

\subsection{Measuring Conversion Factor Distributions} \label{S4.2}
As discussed in many previous atomic carbon studies for more than two decades (see Section~\ref{S1}), extragalactic astronomers have tried to utilize the [\CI]~(1--0) line emission as a tracer of total molecular gas mass by taking advantage of its low to moderate optical depth and the simple three energy level system of the [\CI] lines. However, we do not exactly know the excitation condition and the optical depth of the [\CI]~(1--0) line because of the lack of [\CI]~(2--1) data at the same spatial resolution. Thus, in this section we utilize the optically-thin 609$\mu$m dust continuum as a gold standard molecular gas mass estimator following the prescription described in \citet{Scoville16}, and then calculate conversion factors. Plausibility of the derived conversion factors and a comparison with previous works will be presented in the next section.

\begin{figure*}
\includegraphics[width=2.0\columnwidth]{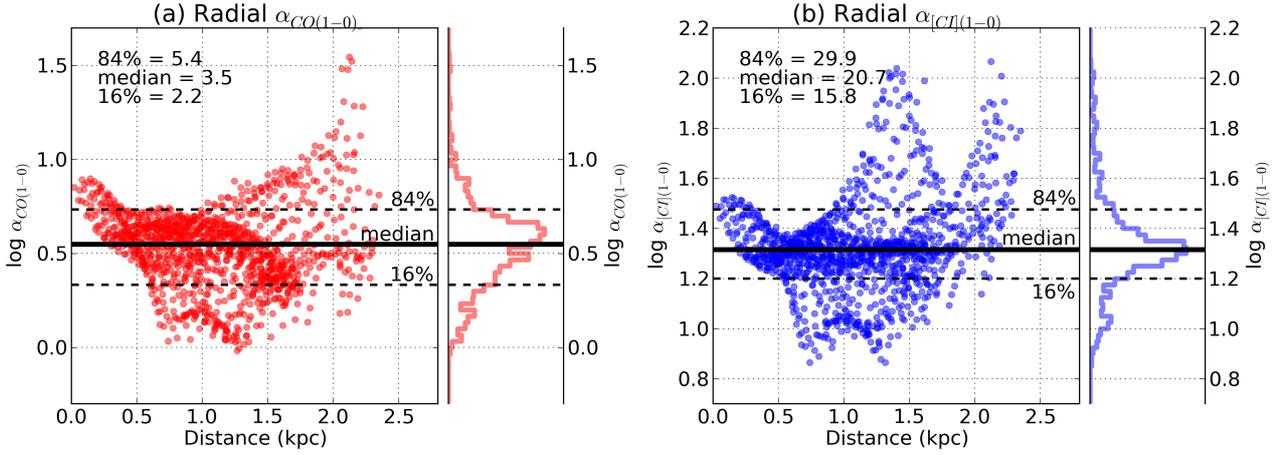}
\caption{
(a) Radial $\alpha_{\rm CO(1-0)}$ (= $M_{\rm H_2}$/$L'_{\rm CO(1-0)}$) distribution and histogram. The solid and two dashed black lines illustrate the 16th, 50th (median), and 84th percentiles.
(b) Same as panel (a), but for $\alpha_{\rm [\CI](1-0)}$ (= $M_{\rm H_2}$/$L'_{\rm [\CI](1-0)}$).
}
\label{fig05}
\end{figure*}

To derive the total molecular gas mass (i.e., H$_2$+He) in units of $M_{\odot}$ we use the equations 6 and 16 of \citet{Scoville16};
\begin{eqnarray} \label{eqn7}
M_{\rm H_2} &=& 1.78 S_{\nu_{\rm obs}}(1+z)^{-4.8}\left(\frac{\nu_{\rm 850\:\mu m}}{\nu_{\rm obs}}\right)^{3.8}D_{\rm L}^2 \notag \\
&& \times \left(\frac{6.7 \times 10^{19}}{\alpha_{850\:\mu m}}\right)\frac{\Gamma_0}{\Gamma_{\rm RJ}}10^{10},
\end{eqnarray}
where $S_{\nu_{\rm obs}}$ is the observed flux densities in units of mJy, $D_{\rm L}$ is the luminosity distance in Gpc, $\alpha_{850}$ is the conversion factor from 850~$\mu$m continuum to total molecular gas mass assuming $T_{\rm dust}$ = 25~K. This is similar to $T_{\rm ex,CO(1-0)}$ and $T_{\rm ex,[\CI](1-0)}$ under typical (U)LIRG conditions \citep{Salak19}. We use $\alpha_{850\:\mu m}$ = 6.7 $\times$ 10$^{19}$ erg s$^{-1}$ Hz$^{-1}$ M$_{\odot}^{-1}$, which is a calibrated value for local star-forming galaxies, (U)LIRGs, and high-z submillimeter galaxies.
$\Gamma_{\rm RJ}$ is given by,
\begin{equation} \label{eqn8}
\Gamma_{\rm RJ}(T_{\rm dust}, \nu_{\rm obs}, z) = \frac{h\nu_{\rm obs}(1+z)/kT_{\rm dust}}{e^{h\nu_{\rm obs}(1+z)/kT_{\rm dust}} - 1},
\end{equation}
and $\Gamma_0$ = $\Gamma_{\rm RJ}(T_{\rm dust}, \nu_{\rm 850\:\mu m}, 0)$. We note that we assume $T_{\rm dust}$ = 25~K.
This is consistent with a grey body fitting to the FIR spectral energy distribution of nearby (U)LIRGs \citep[33.2 $\pm$ 6.2~K;][]{U12}.

Radial distribution of the derived molecular gas masses divided by the CO~(1--0) and [\CI]~(1--0) luminosities, i.e., $\alpha_{\rm CO(1-0)}$ and $\alpha_{\rm [\CI](1-0)}$, respectively, are shown in Figure~\ref{fig05}. The distributions are relatively flat with a peak at the centre. Note that flatter radial $\alpha_{\rm CO}$ distributions are reported in nearby spiral galaxies and the nearby LIRG NGC~1614 \citep{Sandstrom13,Saito17}. The dispersions seen at large radii are due to low signal-to-noise ratios of the dust continuum data, which are already seen in the weighted and unweighted histograms in Figures~\ref{fig04}.

The median $\alpha_{\rm CO(1-0)}$ for IRAS~F18293-3413 is 3.5~$M_{\odot}$ (K km s$^{-1}$ pc$^2$)$^{-1}$.
This $\alpha_{\rm CO(1-0)}$ is between the empirical value for the Milky Way ($\sim$4.3; \citealt{Bolatto13}) and that of nearby (U)LIRGs ($\sim$0.8; \citealt{Downes98}). The 16\% and 84\% percentiles of the $\alpha_{\rm CO(1-0)}$ distribution is 2.2 and 5.4~$M_{\odot}$ (K km s$^{-1}$ pc$^2$)$^{-1}$, respectively, and thus most of the pixels show $\alpha_{\rm CO(1-0)}$ comparable or slightly lower than the Milky Way value.

We found the $\alpha_{\rm [\CI](1-0)}$ distribution is relatively flat across the molecular gas disk of IRAS~F18293-3413 with a median = 20.7~$M_{\odot}$ (K km s$^{-1}$ pc$^2$)$^{-1}$, 16\%  percentile = 15.8~$M_{\odot}$ (K km s$^{-1}$ pc$^2$)$^{-1}$, and 84\%  percentile = 29.9~$M_{\odot}$ (K km s$^{-1}$ pc$^2$)$^{-1}$, which are larger than $\alpha_{\rm CO(1-0)}$ by an order of magnitude. The flat radial $\alpha_{\rm [\CI](1-0)}$ distribution is a natural consequence of the linear relation between [\CI] and dust (Figure~\ref{fig02}b). \cite{Israel15} reported that the global $\alpha_{\rm [\CI](1-0)}$ values are roughly 10 times larger than the global $\alpha_{\rm CO(1-0)}$ for nearby (U)LIRGs, which support our spatially-resolved measurements here.

Conversion factors derived above are based on the gas masses with a fixed $\alpha_{\rm 850\mu m}$ of 6.7 $\times$ 10$^{19}$~erg s$^{-1}$ Hz$^{-1}$ $M_{\odot}$ \citep{Scoville16}.
Here we briefly discuss how the results change when applying variable $\alpha_{\rm 850\mu m}$.
We use a linear relation proposed by \citep[i.e., luminosity-dependent $\alpha_{\rm 850\mu m}$;][]{Hughes17}, which can be written as,
\begin{equation}
\log_{10}\:M_{\rm H_2} = (0.92 \pm 0.02)\log_{10}\:L_{\nu_{\rm 850}} - (17.31 \pm 0.59)
\end{equation}
where $L_{\nu_{\rm 850}}$ is the continuum luminosity at 850~$\mu$m.
We use the exactly same parameters used in Equation~\ref{eqn8} to convert the observed 609~$\mu$m flux to the 850~$\mu$m flux.

The median value is 4.1 $\times$ 10$^{19}$~erg s$^{-1}$ Hz$^{-1}$ $M_{\odot}$, which makes the resultant $\alpha_{\rm CO(1-0)}$ and $\alpha_{\rm [\CI](1-0)}$ larger than values seen in Figure~\ref{fig05} by $\sim$50\%.
The radial trends seen in Figure~\ref{fig05} does not change, because the derived $\alpha_{\rm 850\mu m}$ is almost constant across the disk of IRAS F18293-3413 (maximum and minimum are 3.8 $\times$ 10$^{19}$ and 4.8 $\times$ 10$^{19}$~erg s$^{-1}$ Hz$^{-1}$ $M_{\odot}$, respectively).

\subsection{Physical Properties of the [\CI]~(1--0)-emitting ISM in IRAS F18293-3413}
In this section, we discuss possible physical conditions of the ISM emitting [\CI]~(1--0) line emission which can reproduce the derived $\alpha_{\rm [\CI](1-0)}$ values. We follow the optically-thin prescription described in \citet{Papadopoulos04b} \citep[see also][]{Alaghband-Zadeh13}.

To derive the total molecular gas mass using the [\CI]~(1--0) line, we use the equation;
\begin{eqnarray} \label{eqn9}
M_{\rm H_2} &=& 1375.8 \frac{D_{\rm L}^2}{1 + z} \left(\frac{X_{\rm \CI}}{10^{-5}}\right)^{-1} \notag \\
&& \times \left(\frac{A_{10}}{10^{-7}s^{-1}}\right)^{-1} Q_{10}^{-1} \frac{S_{\rm line}\Delta v}{\rm Jy\:km\:s^{-1}}
\end{eqnarray}
where, $A_{\rm 10}$ is the Einstein coefficient (= 7.93 $\times$ 10$^{-8}$ s$^{-1}$), $X_{\rm \CI}$ is the \CI\, abundance relative to H$_2$, and $Q_{10}$ is the [\CI]~(1--0) excitation factor.

By substituting equation~\ref{eqn1} for equation~\ref{eqn9}, $\alpha_{\rm [\CI](1-0)}$ can be expressed as,
\begin{equation} \label{eqn10}
\alpha_{\rm [\CI](1-0)} \equiv \frac{M_{\rm H_2}}{L'_{\rm [\CI](1-0)}} = 3.357\nu_{\rm obs}^2(1 + z)^2X_{\rm \CI}^{-1}Q_{10}^{-1},
\end{equation}
where $X_{\rm \CI}$ is the abundance of atomic carbon relative to H$_2$ and $Q_{10}$ is the [\CI]~(1--0) partition function. According to the equation, $\alpha_{\rm [\CI](1-0)}$ is simply inversely proportional to $Q_{\rm 10}$ and $X_{\rm \CI}$, while both cannot be constrained by the current dataset. Thus, we explore $Q_{\rm ul}$ and $X_{\rm \CI}$ values in order to reproduce the derived $\alpha_{\rm [\CI](1-0)}$.

\begin{figure}
\includegraphics[width=\columnwidth]{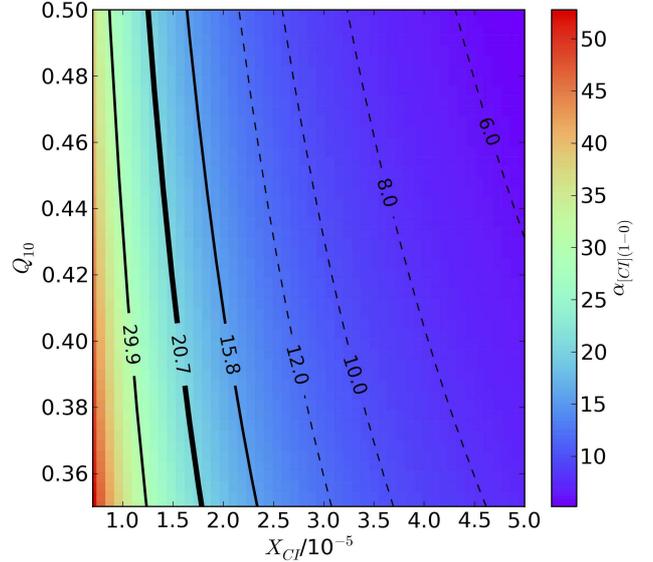}
\caption{
The $\alpha_{\rm [\CI](1-0)}$ distribution in the $Q_{\rm 10}$--$X_{\rm \CI}$ plane.
The solid lines highlight the 16\%, 50\%, and 84\% percentiles of the derived $\alpha_{\rm [\CI](1-0)}$ distribution for IRAS F18293-3413.
The dashed lines highlight other $\alpha_{\rm [\CI](1-0)}$ values for reference.
}
\label{fig06}
\end{figure}

In Figure~\ref{fig06}, we plot $\alpha_{\rm [\CI](1-0)}$ values within reasonable ranges of $Q_{\rm 10}$ and $X_{\rm \CI}$. We found that the derived $\alpha_{\rm [\CI](1-0)}$ for IRAS F18293-3413 can be reproduced within $Q_{\rm 10}$ = 0.35--0.50, and $X_{\rm \CI}$ = (0.8--2.3) $\times$ 10$^{-5}$.

In the case of optically-thin [\CI] emission and weak radiation limit (i.e., $T_{\rm kin}$ $\gg$ background temperature $\sim$ 2.73~K), we can estimate $Q_{\rm 10}(T_{\rm kin},n)$ by using Equations (A21) and (A25) of \citet{Papadopoulos04b}. Based on a non-LTE $Q_{\rm 10}$ grid in the $n$--$T_{\rm kin}$ plane presented in Figure 3c in \citet{Jiao17}, $Q_{\rm 10}$ varies from 0.35 to 0.50 for the typical molecular gas conditions observed in nearby (U)LIRGs ($n$ = 10$^3$-10$^4$~cm$^{-3}$ and $T_{\rm kin}$ = 20-50~K). From Figure~\ref{fig06}, under the non-LTE assumption, the typical molecular gas conditions in nearby (U)LIRGs naturally explain the derived $\alpha_{\rm [\CI](1-0)}$ when $X_{\rm \CI}$ = (0.8--2.3) $\times$ 10$^{-5}$.

$X_{\rm \CI}$ is observationally known to vary from 4-8 $\times$ 10$^{-5}$ \citep[high-redshift SMGs and quasars;][]{Alaghband-Zadeh13,Walter11} to 4 $\times$ 10$^{-5}$ \citep[nearby galaxies;][]{Israel01,Israel03}, and to 2 $\times$ 10$^{-5}$ \citep[Galactic star-forming regions;][]{Frerking89} with large systematic uncertainties depending on the method to estimate the H$_{2}$ column density. Recently, \citet{Valentino18} reported 1.6-1.9 $\times$ 10$^{-5}$ for main-sequence galaxies at z = 1-2, which is consistent with or lower than the previous studies. There seems to be a trend that active objects show higher $X_{\rm \CI}$. The $\alpha_{\rm [\CI](1-0)}$ value decreases as $X_{\rm \CI}$ increases with a negligible relative influence of $Q_{10}$ (Figure~\ref{fig06}), so that active objects showing higher $X_{\rm \CI}$ imply lower $\alpha_{\rm [\CI](1-0)}$ ($\sim$5--10~$M_{\odot}$ (K km s$^{-1}$ pc$^2$)$^{-1}$). This is the same trend as for $\alpha_{\rm CO(1-0)}$ which shows a systematic variation from the Milky Way ($\sim$4.3~$M_{\odot}$ (K km s$^{-1}$ pc$^2$)$^{-1}$) to (U)LIRGs ($\sim$0.8~$M_{\odot}$ (K km s$^{-1}$ pc$^2$)$^{-1}$).

\citet{Heintz20} found that $\alpha_{\rm [\CI](1-0)}$ monotonically decrease as the metallicity increases in absorption-selected high-redshift galaxies. Since $X_{\rm \CI}$ is considered to be mainly driven by metallicity \citep{Glover16}, this study supports the expected relation between $\alpha_{\rm [\CI](1-0)}$ and $X_{\rm \CI}$ (see also \citealt{Crocker19}).

Moreover, one may expect that more intense star-forming environments show higher $X_{\rm \CI}$ due to large cosmic ray (CR) flux and strong UV radiation \citep{Papadopoulos04b}.
This suggests a possible CR-driven radial $X_{\rm \CI}$ distribution in the disk of galaxies, leading to lower $\alpha_{\rm [\CI](1-0)}$ for the centre of IRAS F18293-3413. Thus, the radial $\alpha_{\rm [\CI](1-0)}$ distribution may become flatter, once one can estimate the impact of CR on the spatial $X_{\rm C_I}$ distribution.

The assumed [\CI] line physical parameters significantly affect spatially-resolved molecular gas mass measurements as briefly described above. In order to constrain both $Q_{\rm 10}$ and $X_{\rm C_I}$ distributions, multi-CO radiative transfer analysis and the higher fine structure of atomic carbon line, [\CI]~(2--1), are needed.

\section{Conclusions} \label{S5}
In this paper, we present high-quality ALMA observations of [\CI]~(1--0) CO~(1--0), and 609~$\mu$m dust continuum emission toward the nearby LIRG IRAS~F18293-3413 at 300-pc ($\sim$0\farcs8) resolution, providing spatially-resolved measurements of both lines and the continuum for the whole gas disk of a single galaxy with sufficient resolution and sensitivity.
We detect both lines with significant signal-to-noise to characterize their spatial distributions.
The [\CI]/CO line luminosity ratio is complete above a CO~(1--0) luminosity = 10$^6$~K km s$^{-1}$ pc$^2$, while it is sufficient to measure radial trends up to 2.5~kpc distance from the centre.
Within this distance, we also detect the continuum emission.

We find that the [\CI]~(1--0) line traces structures detected in CO~(1--0) and 609~$\mu$m dust emission, and the [\CI] and dust flux distributions are apparently comparable or slightly more compact than CO~(1--0).
However, a pixel-by-pixel comparison revealed that there is a superlinear relation (slope = 1.54 $\pm$ 0.02) between \CI\, and CO luminosities
and a linear relation between \CI\:and 609~$\mu$m dust luminosities.
Those relations can be explained by radially varying excitation temperature and/or line opacity gradients.

The [\CI]~(1--0)/CO~(1--0) ratio histogram is well fitted by a single Gaussian component with $\mu$ = 0.16 and $\sigma$ = 0.04, and the scaled [\CI]~(1--0)/dust ratio histogram shows a similar distribution.
However, the luminosity-weighted [\CI]~(1--0)/CO~(1--0) ratio histogram shows an offset from a Gaussian due to the high line ratios around the centre of IRAS~F18293-3413.

We convert the 609~$\mu$m dust emission to the molecular gas mass in order to calculate radial distributions of the CO~(1--0)-to-H$_2$ ($\alpha_{\rm CO(1-0)}$) and [\CI]~(1--0)-to-H$_2$ ($\alpha_{\rm [CI](1-0)}$) conversion factors within the disk of IRAS~F18293-3413.
The derived conversion factors are relatively constant values (median $\alpha_{\rm CO(1-0)}$ = 3.5 and $\alpha_{\rm [CI](1-0)}$ = 20.7~$M_{\odot}$ (K km s$^{-1}$ pc$^2$)$^{-1}$) across the disk with some dispersions at larger distance from the centre which are mainly driven by low signal-to-noise data points.
A non-LTE calculation yields that typical molecular gas properties ($n_{\rm H_2}$ = 10$^{3-4}$~cm$^{-3}$, $T_{\rm kin}$ = 40-50~K, and $X_{\rm\CI}$ = (0.8-2.3) $\times$ 10$^{-5}$) can naturally explain the derived $\alpha_{\rm [CI](1-0)}$.

The [\CI]~(1--0) line can be used to trace the molecular gas kinematics and infer flux distribution of CO~(1--0) and dust emission within the gas disk of galaxies.
However, a careful treatment on the gas physical properties (e.g., radial slope of gas temperature and \CI\, abundance) is required in order to measure H$_2$ gas mass distribution in galaxies using a single [\CI]~line.
Otherwise, a [\CI]~line is not a good molecular gas estimator in a spatially resolved manner.

\section*{Acknowledgements}
The authors thanks an anonymous referee for comments that significantly improved the contents of this paper.
T.S., D.L. and E.S. acknowledge funding from the European Research Council (ERC) under the European Union's  Horizon 2020 research and innovation programme (grant agreement No. 694343).
Y.A. acknowledges financial support by NSFC grant 11933011.
This paper makes use of the following ALMA data: ADS/JAO.ALMA$\#$2015.1.01191.S and ADS/JAO.ALMA$\#$2018.1.00994.S.
ALMA is a partnership of ESO (representing its member states), NSF (USA) and NINS (Japan), together with NRC (Canada), MOST and ASIAA (Taiwan), and KASI (Republic of Korea), in cooperation with the Republic of Chile. The Joint ALMA Observatory is operated by ESO, AUI/NRAO and NAOJ.
This research has made use of the NASA/IPAC Extragalactic Database (NED) which is operated by the Jet Propulsion Laboratory, California Institute of Technology, under contract with the National Aeronautics and Space Administration.
This research made use of Astropy, a community-developed core Python package for Astronomy \citep{Astropy13,Astropy18}.

\section*{Data Availability}

The data underlying this article were accessed from the ALMA science archive \footnote{\url{http://almascience.eso.org/aq/}}. The derived data generated in this research will be shared on reasonable request to the corresponding author.






\bsp	
\label{lastpage}
\end{document}